\documentstyle[preprint,aps,prd,epsfig,floats]{revtex}

\begin{document}

\tighten

\preprint{\parbox[t]{50mm}
{\begin{flushright}
ADP-00-14/T398 \protect\\ 
PSI-PR-00-05 \protect\\
UCY-PHY-00/01 
\end{flushright}}}

\vspace{2cm}

\title{Non-Perturbative Mass Renormalization in Quenched QED \protect \\
from the Worldline Variational Approach}

\vspace{1cm}
 
\author{C.~Alexandrou $^1$, R.~Rosenfelder $^2$ and A.~W.~Schreiber $^3$}

\vspace{1cm}

\address{$^1$ Department of Physics, University of Cyprus, 
              CY-1678 Nicosia, Cyprus\\
         $^2$ Paul Scherrer Institute, CH-5232 Villigen PSI, Switzerland\\
         $^3$ Department of Physics and Mathematical Physics and
              Research Centre for the Subatomic Structure of Matter,
              University of Adelaide, Adelaide, S. A. 5005, Australia}

\date{June 14, 2000}

\maketitle

\begin{abstract}
Following Feynman's successful treatment of the polaron problem we
apply the same variational principle to quenched QED in the worldline
formulation. New features arise from the description of fermions by
Grassmann trajectories, the supersymmetry between bosonic and
fermionic variables and the much more singular structure of a
renormalizable gauge theory like QED in $3 + 1$ dimensions.  We take
as trial action a general retarded quadratic action both for the
bosonic and fermionic degrees of freedom and derive the variational
equations for the corresponding retardation functions. We find a
simple analytic, non-perturbative, solution for the anomalous mass
dimension $\gamma_m(\alpha)$ in the MS scheme. For small couplings we compare
our result with recent four-loop perturbative calculations while at
large couplings we find that $\gamma_m(\alpha)$ becomes proportional
to $\sqrt \alpha$.  The anomalous mass dimension shows no obvious sign
of the chiral symmetry breaking observed in calculations based on the
use of Dyson-Schwinger equations, however we find that a perturbative
expansion of $\gamma_m(\alpha)$ diverges for $\alpha > 0.7934$.
Finally, we investigate the behaviour of $\gamma_m(\alpha)$ at large orders in
perturbation theory.
\end{abstract}


\section{Introduction}

Variational methods are widely used in many areas of physics but are not 
very prominent in field theory \cite{var}. This is due to the infinite number 
of degrees of freedom and the singular short-distance behaviour
of relativistic field theories. A very successful application of
variational methods in a {\it non-relativistic}
field theory is provided by Feynman's treatment of the polaron \cite{Fey}: 
after integrating out the phonon degrees of freedom and 
approximating  variationally the remaining effective action by a retarded 
quadratic trial action one obtains the best approximation scheme which works 
for both small and large coupling constants. Detailed numerical investigations
\cite{AlRo} have shown that Feynman's approximate solution deviates at most 
$ 2.2 \%$ from the true ground state energy for all coupling constants. 
It is therefore very attractive
to apply similar techniques to problems in relativistic quantum field theory 
where there is much need for non-perturbative methods. In previous publications
we have done that in the context of a scalar, super-renormalizable
model theory \cite{WC}.

\vspace{0.1cm}

In this paper we present the first results obtained by applying
polaron variational methods to a realistic theory, namely Quantum
Electrodynamics (QED) in the quenched approximation where
electron-positron loops are neglected.  While the actual coupling
constant between electrons and photons $ \alpha = e^2/(4 \pi) \simeq
1/137$ is small enough to apply perturbation theory in most cases,
there is enough interest to study the theory at larger coupling:
first, the strong coupling behaviour of any physical theory is of
interest in itself, second, the possibility of chiral symmetry
breaking \cite{chiSBRev} demands an investigation at large $\alpha$ and,
finally, bound state problems are inherently non-perturbative and
involve powers of $\ln 1/\alpha \simeq 4.92$ in radiative corrections.

\vspace{0.1cm}

The extension of our methods to QED requires a formalism to include
fermions and a treatment of the more severe singularities encountered
in a renormalizable field theory rather than a super-renormalizable or
non-relativistic one. We do this within the worldline technique which
has recently experienced a revival\cite{worldline}. In this
formulation, the degrees of freedom describing the electron are its
bosonic worldline $x_{\mu}(t)$, which is the four-dimensional analogue
to the polaron trajectory, as well as a Grassmannian path
$\zeta_{\mu}(t)$ needed to describe the electron's spin
\cite{remark1}.  Here $t$ is the proper time which parametrizes the
paths and runs from $0$ to $T$. 
The dynamics of the electron in an external vector field $A_{\mu}(x)$ 
with field strength $F_{\mu \nu}(x) $ are then described by the following 
worldline Lagrangian
\begin{equation}
L \>=\> -\frac{\kappa_0}{2} \dot{x}^2 + i \zeta \cdot \dot \zeta +
\,\frac{1}{T} \,\dot x \cdot \zeta \, \chi - e \,\dot{x} \cdot A(x) -
\frac{ie}{\kappa_0} F_{\mu\nu} (x) \, \zeta^{\mu}\zeta^{\nu} \> .
\end{equation}
Here $\kappa_0$ is an arbitrary parameter which may be used to
reparametrize the
proper time without changing the physics and $\chi$ is a Grassmannian
(super-)partner of the proper time $T$.  Note that the above action exhibits 
a well-known
{\it supersymmetry} between bosonic and fermionic degrees of freedom
\cite{BDZVH}. For further details about the application of the worldline 
formalism to QED 
we refer the reader to Ref.~\cite{QED1}.

The photon field $A_{\mu}$ may be integrated out exactly in complete 
analogy to the phonons in the polaron case, resulting in an
effective action for the electron only
\begin{eqnarray}
S_{\rm eff} &=& S_0 - \frac{e^2}{2} \int_0^T dt_1 \, dt_2 \int
\frac{d^4 k}{(2 \pi)^4} \> G^{\mu\nu}(k)  \>
\left [ \, \dot x_{\mu}(t_1) \, + \,
\frac{2}{\kappa_0} \zeta_{\mu}(t_1) \, k \cdot \zeta(t_1) \, \right ]
\nonumber \\
&&  \hspace{3cm} \cdot \left [ \, \dot x_{\nu}(t_2) \, - \,
\frac{2}{\kappa_0} \zeta_{\nu}(t_2) \> k \cdot \zeta(t_2) \, \right ]
\> e^{- i k \cdot \left [ \, x(t_1) - x(t_2) \, \right ] } \> .
\label{S eff}
\end{eqnarray}
Here $S_0$ denotes the free action 
\begin{equation}
S_0 \> = \> \int_0^T \> dt \>\left [-\frac{\kappa_0}{2} 
\dot{x}^2(t) + i \zeta(t) \cdot \dot \zeta(t) +
\,\frac{1}{T} \,\dot x(t) \cdot \zeta(t) \, \chi \right ]
\label{S0}
\end{equation}
and $ G^{\mu\nu}(k)$ the
gauge-fixed photon propagator.  As described in Ref.~\cite{QED1},
the electron propagator is obtained
by carrying out a path integral over the degrees of freedom $x_{\mu}(t)$ and
$\zeta_{\mu}(t)$, as well
as a weighted integral over the proper times $T$ and $\chi$, and finally
identifying the Grassmannian variable $\Gamma = \zeta(0) + \zeta(T)$ 
with the Dirac matrix $\gamma$:
\begin{eqnarray}
G_2(p) & = & e^{\gamma \cdot {\partial \over \partial \Gamma}}
\int_0^\infty dT \> \int d\chi \> \exp \left \{
{i \over 2 \kappa_0} \left [(p^2-M_0^2) T + (p \cdot \Gamma - M_0)\chi \right ]
\right \} \nonumber \\
&& \cdot
\left .\frac{\int {\cal D}  \tilde x \,{\cal D} \zeta \, 
\exp \left [ i p \cdot x + \zeta(0)\cdot \zeta(T) \right ] \, 
\exp\left ( i S_{\rm eff} \right )}{ \int {\cal D} 
\tilde x \, {\cal D} \zeta \, \exp \left [ i p \cdot x + \zeta(0)\cdot 
\zeta(T) \right ]
\exp( i S_0 ) }\right |_{\Gamma=0} \>\;\;.
\label{exact G2}
\end{eqnarray}
Here $M_0$ is the bare mass and ${\cal D}  \tilde x$ contains an
integration over
the endpoint $ x = x(T) $. Note that we have divided 
and multiplied by the path integral for the 
free theory, so the bare propagator may be obtained by just ignoring
the last line.  For non-zero couplings, of course, the path integrals in
the last line cannot be performed;  these  we shall approximate variationally
in the next section.

\section{Variational Approach}

Feynman's variational principle has its root in Jensen's inequality
for convex functions applied to $\exp(-S_E)$, where $S_E$ is a
Euclidean action. In Minkowski space and/or for complex actions 
the variational principle remains valid, however it becomes a stationary 
principle rather than a minimum principle.  To be more precise,
the path integral over bosonic and fermionic paths obeys
\begin{equation}
\Bigl < \, \exp \left [ \, i (S - S_t) \, \right ]  \, \Bigr >_t \>
\stackrel{\rm stat.}{\simeq} \>  
\exp \left [ \,  i \left < S - S_t \right >_t  \, \right ]  \> \;\;\;,
\label{Jensen}
\end{equation}
where $<\ldots>_t$ indicates an average 
involving the weight function $e^{i S_t}$ in the relevant functional 
integral and $S_t$ is a suitable
trial action.  Note that corrections to this variational approximation
may be calculated in a systematic way and that, furthermore, to first order 
in the interaction
(i.e. to order $\alpha$) the relation is in fact an equality if $S_t$ reduces  
to the free action for small couplings.

\vspace{0.1cm}
For the trial action
required in Eq. (\ref{Jensen}) we choose a general
{\it retarded} quadratic action which is a two-time modification of the 
free action in Eq. (\ref{S0}) 
\begin{eqnarray}
\tilde S_t &=& S_0 + i \kappa_0^2 \int_0^T \! dt_1 dt_2 \,  \Biggl [ \, 
- g_B(\sigma) \, \dot x(t_1) \cdot \dot x(t_2) + \frac{2i}{\kappa_0} 
g_F'(\sigma) \, \zeta(t_1) \cdot \zeta(t_2) \nonumber \\ 
&& \hspace{0.5cm} - \, 2 \, \frac{\sigma}{\kappa_0 T} \, g_{SO}'(\sigma) 
\, \dot x(t_1) \cdot \zeta(t_2) \chi \, \Biggr ] \> + \> 
\lambda_1 \, p \cdot  x - i
 \lambda_2 \, \zeta(0) \cdot \zeta(T) \, \>.
\label{S trial}
\end{eqnarray}
Here the variational parameters are contained in the retardation
functions $g_i(\sigma)$ for bosonic, fermionic and spin-orbit
interactions; these are even functions of $\sigma = t_1 - t_2 $ and
they become identical for a supersymmetric trial 
action~\footnote{We have explicitly separated out the free action in 
Eq.~(\protect\ref{S trial}), which could have alternatively been
added into the second term by adding
$\delta(\sigma)/(2i\kappa_0) $  to each of the $g_i(\sigma)$'s.
This way  our retardation functions contain no distributions.
Also, in the supersymmetric limit our trial action could be
written in the explicitly supersymmetric notation of Ref.~\cite{QED1} as
$i \kappa_0^2 \int_0^T dt_1 dt_2 \, \int d\theta_1 d\theta_2 \> 
g(T_{12}) \, DX_1 \cdot DX_2$, with a single retardation function.}.
The variational principle `adjusts' these functions in
order to compensate for the fact that the true effective action
(\ref{S eff}) is not quadratic in the variables $x(t)$,
$\zeta(t)$. Feynman's polaron result was obtained by taking a specific
{\it Ansatz} for the retardation functions but here we leave their
functional form free.  This is because one expects that the correct
short-time behaviour of these functions is much more important for a
renormalizable theory like QED than for the polaron problem which does
not exhibit any ultraviolet divergences. Indeed one finds that for
small $\sigma$ the ``best'' $g_B(\sigma)$ behaves like $\sqrt{\sigma}$,
$\ln \sigma$ and $1/\sigma$ in the polaron, super-renormalizable and
QED case, respectively. The ``tilde'' over the trial action indicates
that it includes the boundary terms already present in Eq.~(\ref{exact G2})
and that
we are using ``momentum averaging'' \cite{WC}. These 
terms, involving the external momentum $p$ and the Grassmann variable
$\Gamma$, are multiplied by additional variational parameters $\lambda_1$ and
$\lambda_2$, respectively. They provide additional freedom to
 modify the strength of the boundary terms.  We have allowed
this freedom because of our experience in scalar relativistic
field theory~\cite{WC}, where the variational parameter $\lambda_1$
turned out to be essential for describing the instability of the
Wick-Cutkosky model.

\vspace{0.1cm} Since the trial action (\ref{S trial}) is at most
quadratic in $x(t)$ and $\zeta(t)$ it is possible to evaluate the
various averages required in Eq. (\ref{Jensen}) analytically.  
A particular simplification occurs if one restricts oneself to
$p^2=M^2$, where $M$ is the physical (i.e. pole) mass:  as 
discussed in \cite{WC}, the divergence of the propagator
on its mass shell results from a divergence of the integral over
the proper time $T$. Indeed, the variational approximation (\ref{Jensen})
results in an electron propagator [see Eq.~(\ref{exact G2})] which has
the form
\begin{eqnarray}
G_2^{\rm var}(p) & = & e^{\gamma \cdot {\partial \over \partial \Gamma}}
\int_0^\infty dT \> \int d\chi \> \exp \left \{
{i T \over 2 \kappa_0} \left [-M_0^2 + 
p^2 ( 2 \lambda - \lambda^2)\right ]\right \}
\label{eq: var g2} \\
&&\hspace{1cm} \cdot \exp \left. \left \{ - {i T \over  \kappa_0} \left (
\> \Omega[A_B] - \Omega[A_F] +  V [\mu^2_B,\mu^2_F]  \> \right )
\>  + \> F(\chi,\Gamma; T;p) \right \}\right |_{\Gamma=0},
\nonumber
\end{eqnarray}
where $\lambda$ (which is defined below), 
 the $\Omega$'s and 
$V$ are $T$-independent and
the function $ F(\chi,\Gamma; T;p)$ is subleading in $T$.  The latter
therefore contains information relevant for the wavefunction
renormalization of $G_2^{\rm var}(p)$, and not the pole structure.
We leave the 
discussion of this function,
which also contains the entire $\chi$ and $\Gamma$ dependence (and hence
the spin structure of the propagator),
for a future publication as it is not required for 
our present investigation.

From Eq.~(\ref{eq: var g2}) we see that the bare and physical mass are related
through
\begin{equation}
M_0^2 \> = M^2 ( 2 \lambda - \lambda^2) - 2 \Biggl ( \> \Omega[A_B] 
- \Omega[A_F]  + V [\mu^2_B,\mu^2_F]  \> \Biggr ) \> .
\label{Mano eq}
\end{equation}
We have labeled this relationship {\it Mano's
equation} as K. Mano first applied polaron techniques to a scalar
relativistic field theory \cite{Mano}.  Note that, on mass shell, the
variational equations resulting from Eq.~(\ref{Jensen}) are equivalent
to demanding stationarity of Mano's equation.

The nomenclature in Mano's equation corresponds to that introduced in
Ref.~\cite{WC}:  $\Omega[A_B]$ and $\Omega[A_F]$ originate from contributions 
(bosonic and fermionic, respectively) of the terms in Eq.~(\ref{Jensen}) 
involving
$S_0$ and $S_t$ only.  They are the analogue to the kinetic term in variational
quantum mechanical calculations, while the analogue of the contribution from a
potential term (explicitly proportional to the strength of the coupling)
resides in $V$.

Similarly to Ref.~\cite{WC}, it is useful to express the retardation functions in
terms of
the variational ``profile functions'' $A_i(E)$ and the ``pseudotimes'' 
$\mu^2_i (\sigma) \>, i = B,F$ defined by
\begin{eqnarray}
A_i(E) &=& 1 + i \kappa_0 \int_0^{\infty} d\sigma \,
 g_i(\sigma) \cos(E \sigma) 
\label{def profile func}\\
\mu^2_i (\sigma) &=& \frac{4}{\pi} \int_0^{\infty} dE \> \frac{1}{E^2 A_i(E)}
\, \sin^2 \left ( \frac{E \sigma}{2} \right ) \> ,
\label{def pseudotime}
\end{eqnarray}
respectively~\footnote{It is easy to show from Eq.~(\ref{def pseudotime}) 
that, as
long as $A_i(0)$ and $A_i(\infty)$ exist, for both asymptotically large and 
infinitesimally small times $\sigma$ 
the functions
$\mu^2_i (\sigma)$ become proportional to $\sigma$; hence the label
 ``pseudotime''.  In the free case one has 
$A_i(E)  = 1, \mu_i^2(\sigma) = \sigma, \lambda = 1, \Omega_i = 0 $. The other
variational function $A_{SO}(E)$ and parameter $\lambda_2$ are
 linked to the spin structure of the 
propagator
and therefore do not show up in Mano's  equation.}.
Furthermore, it is convenient to define $ \lambda = \lambda_1/A_B(0)$. 
Indeed it turns out that the  
averages in Eq.~(\ref{Jensen}) can be directly expressed in terms of these 
quantities.
The kinetic terms in Eqs.~(\ref{eq: var g2}) and~(\ref{Mano eq}) become
\begin{equation}
\Omega[A_i] \> = \> {d \>  \kappa_0 \over 2 i \pi} \int_0^\infty \> dE \>
\left ( \log A_i(E) \> + \> {1 \over A_i(E)}\> - \> 1 \right )\;\;\;,
\label{omega_i}
\end{equation}
where $d$ is the spacetime dimension $d = 4 - 2\epsilon$. This is identical 
to the result in Ref.~\cite{WC} if $d = 4$ and $\kappa_0 = i$ (i.e. the
Euclidean  formulation) are taken. 
The  specific properties of QED are encoded in 
the ``interaction'' term
$V$ which, with $V = V_1 \> + \> V_2$,  reads
\begin{eqnarray}
V_1 \, [\mu_B^2,\mu^2_F] &=& - (d-1) \pi \alpha \, 
\frac{\nu^{2 \epsilon}}{\kappa_0} \,
\int_0^{\infty} d\sigma \> \int \frac{d^dk}{(2 \pi)^d} \>
\left \{ \,\left [\dot \mu^2_F(\sigma)\right ]^2 - 
\left [\dot \mu^2_B(\sigma)\right ]^2 \> \right \} \, 
E(k,\sigma) 
\label{V1} \\
V_2 \, [\mu^2_B] &=& - \frac{4 \pi \alpha \> \nu^{2 \epsilon} \lambda^2}
{\kappa_0} \int_0^{\infty} d\sigma  \int \frac{d^dk}{(2 \pi)^d} \> 
\frac{1}{k^2} \left [  \, M^2
+ (d-2) \frac{ (k \cdot p)^2}{k^2} \, \right ] \, E(k,\sigma)\;\;\;.
\label{V2}
\end{eqnarray}
Note that by $\dot \mu^2(\sigma)$ we mean ${d \over d \sigma} \mu^2(\sigma)$
(and not $\left [{d \over d \sigma} \mu(\sigma)\right ]^2$), the
function $ E(k,\sigma)$ is defined to be
$ E(k,\sigma) = \exp \left \{ i \left [ k^2 \mu^2_B(\sigma) - 2
\lambda k \cdot p \sigma \right ]/(2 \kappa_0) \right \} $ and of course $p^2 =
M^2$. The fermionic contributions, both in the `kinetic
term' $\Omega_F$ as well as in $V_1$, appear  with an opposite sign to
the bosonic contributions.  The reason for the separation of $V$ into 
two pieces will become apparent below.  

\vspace{0.1cm}

By construction Mano's equation is {\it stationary} under variation of
the parameters. It is important to note that we have not demanded the
various retardation functions $g_{B,F}$ (as well as $g_{SO}$, which only plays 
a role for the residue)  to be 
identical (before
variation).  Had we done so, the resulting profile functions $A_B$ and
$A_F$ would have also been identical, the pseudotimes $\mu^2_{B,F}$
would have been one and the same and hence $\Omega[A_B]-\Omega[A_F]$
as well as $V_1$ would have vanished.  The absence of a `kinetic'
contribution would have been fatal to the variational principle as
this contribution provides the restoring `force' to the potential $V$.
On the other hand, closer examination of $V_1$ reveals that
$\dot\mu^2_{B} \ne \dot\mu^2_{F}$ is also dangerous: The contribution
of each of these terms is quadratically (UV) divergent if the
dimensional regularization is replaced by a momentum cutoff.  This may be 
checked by either directly substituting the small $\sigma$ limit of 
$\mu^2_{i}(\sigma)$
into $V_1$ or by noting that for scalar QED, where the Grassmannian
path integrals are absent, the remaining contribution from 
$\dot\mu^2_{B}$ gives rise to the quadratically divergent one-loop diagram of that
theory.  It is {\it the combination} $(\dot\mu^2_{B})^2 - (\dot\mu^2_{F})^2$ 
which displays the usual logarithmic UV
divergence of QED.  Although at leading order in the coupling we are
guaranteed to reproduce the correct perturbative result [see
Eq.~(\ref{Jensen})], at higher orders the cancellation of these
quadratic divergences is ensured by the supersymmetry.  To summarize,
on the one hand the trial action cannot be restricted to contain only
supersymmetric terms but on the other hand allowing non-supersymmetric
terms may destroy the renormalizability of the theory.

The way out of this predicament is provided by the
variational principle itself: although it is unavoidable that the
trial action breaks supersymmetry, the actual solutions to the
variational equations may in fact be nearly supersymmetric.  That
this indeed turns out to be the case may be seen by recognizing that
$V_1$ is the most singular part of the interaction whereas $V_2$,
which involves only bosonic contributions and is the only source of
supersymmetry breaking, is similar in structure to the scalar
super-renormalizable model studied before \cite{remark5}. Divergent
contributions in the limit $\epsilon \to 0$ to the variational
equations are solely determined by $V_1$. Therefore, the divergent
contributions to $A_B(E)$ and $A_F(E)$, and hence
to $\dot \mu_B^2$ and  $\dot \mu_F^2$, are identical. 

In this paper we confine ourselves to studying this divergent structure
and so it is sufficient to set $ A_B(E) = A_F(E) \equiv A(E) $.
The corresponding variational equation becomes, after performing the 
$k$-integration in Eq. (\ref{V1}),
\begin{equation}
A(E) \>=\> 1 + (1-\epsilon) \, c_{\epsilon} \, \nu^{2 \epsilon}
\, \int_0^{\infty} d\sigma \> \frac{\sin E\sigma}{E} \,
\frac{\dot \mu^2(\sigma)}{\left [ \mu^2(\sigma) 
\right ]^{2-\epsilon}} \, \exp \left [ -i 
\frac{\lambda^2 M^2 \sigma^2}{2 \kappa_0 \mu^2(\sigma)} \right ] \;\;\;.
\label{var eq for A(E)}
\end{equation}
Note that here we have now also dropped the subscript on the
pseudotime as it is no longer relevant and we have defined
\begin{equation}
 c_{\epsilon}\> = \> {\alpha \over \pi} 
\left ( {2 \pi i \over \kappa_0} \right )^{\epsilon} 
\, {3- 2 \epsilon \over (1-\epsilon)(2 - \epsilon) } 
\> \stackrel{\epsilon \to 0}{\longrightarrow} \> 
\frac{3 \alpha}{2 \pi} \> .
\end{equation}
 Since
$\mu^2(\sigma) \to \sigma$ for small $\sigma$ one sees that the
$\sigma$-integral in Eq. (\ref{var eq for A(E)}) would diverge for
$\epsilon = 0$; this just reflects the $1/\sigma$ behaviour of the
retardation function in Eq.~(\ref{def profile func}) as was discussed
before.  The crucial difference between super-renormalizable and
renormalizable theories therefore is that for the latter ones the
variational equations themselves are UV-divergent. In this way the
divergent structure of higher-order diagrams is effectively summed
up. 

We may now simplify $V$ by making use of the above ``asymptotic''
supersymmetry.  The only remaining contribution is that of 
$V_2\equiv \lambda^2 M^2 W_2$
which becomes, after carrying out the integration over the momentum
$k$,
\begin{equation}
W_2\>=\> {(2-\epsilon)(1 - \epsilon) \over 2}
 c_\epsilon \, \nu^{2 \epsilon}
\int_0^\infty {d\sigma  \over \left [\mu^2(\sigma)\right ]^{1 - \epsilon}}
\int_0^1 { du \over  u^\epsilon} 
\> [\epsilon + (1-\epsilon) u]
\exp \left ( - {i \over 2 \kappa_0} 
{\lambda^2 M^2 \sigma^2 \over \mu^2(\sigma)} u
\right )
\label{V2 integrated}
\end{equation}
where the $u$ integration arises from an exponentiation of the
photon propagator in Eq.~(\ref{V2}) in a similar way as in Ref.~\cite{WC}.
With this, the variational equation for $\lambda$ in this asymptotic
limit becomes
\begin{equation}
\lambda \> = \> 1 - {\partial \over \partial \lambda} (\lambda^2 W_2)\;\;\;.
\label{eq: var eq for lambda}
\end{equation}

\section{Mass Renormalization}

Renormalizability of (quenched) QED means that all divergences
can be collected in the mass and wave function renormalization
constants. In the present investigation we concentrate on the mass
renormalization constant in the MS scheme, $Z_M^{\rm MS}$, defined 
via $ M_0 = Z_M^{\rm MS} M_{\nu}$ where
$M_{\nu}$ is an intermediate mass scale. In this scheme it
has the perturbative expansion
\begin{equation}
 Z_M^{\rm MS} \>=\> 1 +  \frac{b_{11}}{\epsilon} \, \frac{\alpha}{\pi} +  
\left [ \frac{b_{22}}{\epsilon^2} +  \frac{b_{12}}{\epsilon} \right ] \,
 \left ( \frac{\alpha}{\pi} \right )^2
+ \ldots \> \> ,
\end{equation}
where it is known from perturbation theory~\cite{tHo} that the
expansion coefficients $b_{ij}$ are pure, i.e. mass independent, numbers.
Furthermore, the renormalization group provides relations between
many of these coefficients;  at order $n$ in perturbation theory
only the coefficient $b_{1n}$ contains new information.  This is
encapsulated in the solution of the renormalization group
equation for $Z_M^{\rm MS}$, namely
\begin{equation}
Z_M^{\rm MS}  \>=\> \exp \left [ - \frac{1}{2 \epsilon} \int_0^{\alpha} dx \, 
\frac{\gamma_m (x)}{x} 
\right] \>=\> \exp \left [ - \frac{1}{2 \epsilon} \sum_{n=1}^{\infty} 
\frac{\gamma_{n-1}}{n} \, \left (\frac{\alpha}{\pi} \right )^n \right ] \> ,
\label{gamma_m def}
\end{equation}
where $\gamma_m(\alpha) $ is the anomalous mass dimension of the 
electron~\cite{Coq}. In perturbation theory, $\gamma_m(\alpha)$
can be extracted from perturbative QCD calculations, which have been
performed up to 4-loop order.  One obtains $\gamma_0 = 3/2, 
\gamma_1 = 3/16$~\cite{Tar}, $\gamma_2 = \frac{129}{64} = 2.0156 $
and  $\gamma_3 
=- \frac{1}{128} \left [ \frac{1261}{8} + 336 \, \zeta(3) \right ] =
-4.3868$~\cite{alpha4}.

As the variational calculation is applicable for arbitrary values of the
coupling, comparison to perturbation theory provides a useful guide to
its utility.  As mentioned before, to first order in the coupling 
the calculation is guaranteed to be exact as long as one has used a   
trial action which can reduce to the free action in the limit $\alpha \to 0$.
A genuine test of the variational scheme is only obtained by 
comparing the coefficients in higher order. It should be noted 
that this test is much more demanding
than in the polaron case where one can only compare the numerical value 
of the second-order coefficient for the energy: here, in addition, one 
tests the $\epsilon$-dependence of this coefficient and also whether it is 
mass-independent as it should be in the exact theory.

In order to know $\Omega$ and $V$ at second order in $\alpha$ one
requires the variational parameters up to first order in $\alpha$.
These may be obtained by inserting the zeroth order results 
$ \mu^2(\sigma) = \sigma$ and $\lambda=1$ into the variational equations 
for the profile function (\ref{var eq for A(E)})  and $\lambda$
(\ref{eq: var eq for lambda}).
The solutions then need to be substituted back 
into $\Omega$ [Eq.~(\ref{omega_i})] and $V_2$ [Eq.~(\ref{V2 integrated})]. 
Having done this, $Z_M^{\rm MS}$ may then be extracted from Mano's equation,
yielding  $b_{22}^{\rm var} = 9/32$, which is correct, 
and 
$b_{12}^{\rm var} = 0$, which should be compared to the exact value of
$b_{12} = -3/64$. As in the Wick-Cutkosky model, the
$\lambda$-variation is of crucial importance: for example, fixing 
$\lambda = 1$ would give a wrong result for $b_{22}$ and a logarithmic 
mass-dependence for $b_{12}$.

It is possible to develop the perturbative expansion of the variational
result further, with the result that no mass dependence in the
coefficients appears even at higher order.  Indeed, it turns out that it
is in fact possible to obtain the {\it full} analytic expression for
the anomalous mass dimension in the worldline variational approximation.
We shall sketch the derivation below, leaving the technical details
for the Appendix to this paper.

To begin with, we first drop the mass term in the variational
equation (\ref{var eq for A(E)}) since it only affects long-distance
physics and not the ultra-violet behaviour contained in $Z_M^{\rm MS}$ . 
Then we change variables from $\sigma$ to 
$y = c_{\epsilon} (\nu^2 \sigma)^{\epsilon}$, and equivalently for $E$.
This has the effect of making the system of integral equations 
(\ref{def pseudotime},\ref{var eq for A(E)},\ref{V2 integrated}) 
independent of the coupling.  We write these explicitly
in the Appendix, where it is shown that if, for small $\epsilon$,  the 
pseudotime
has the form
\begin{equation}
{\mu^2(\sigma) \over \sigma} = \exp \left [ - {\omega_0(y)\over \epsilon}
+ O(\epsilon^0)\right ]
\end{equation}
then the anomalous mass dimension may be written in terms of
this function $\omega_0(y)$, i.e. 
\begin{equation}
\gamma_m^{\rm var} \> = \> { v(y_0) \over 1 -  v(y_0)}\;\;\;,
\label{eq: gamma_m in v}
\end{equation}
where $v(y) = y \; \omega'_0(y)$ and $y_0$ is determined by the implicit 
equation
\begin{equation} 
y_0 = {3 \alpha \over 2 \pi} \, e^{ \omega_0(y_0)}\>  .
\label{implicit}
\end{equation}

On the other hand, it can also be shown (see the Appendix)
that the variational equation for the 
pseudotime translates into an equation for $\omega_0(y)$, i.e.
\begin{equation}
{e^{\omega_0(y)} \over y} \> = \> {\pi \over 2} [ \, 1-v(y) \, ] 
 \cot \left [ {\pi \over 2} v(y) \right ] \;\;.
\label{eq: final var eq}
\end{equation}
This equation must in general be solved numerically.  
However, we note that the calculation of the anomalous mass dimension in 
Eq.~(\ref{eq: gamma_m in v}) only requires knowledge of the function $v(y)$ 
at $y=y_0$.  Furthermore, it is remarkable that, at
this value of $y$, the combination  $e^{\omega_0(y)}/y$ is precisely 
the combination that is fixed in terms of the coupling constant 
[see Eq.~(\ref{implicit})].  Hence, at $y=y_0$, the
the L.H.S of Eq.~(\ref{eq: final var eq}) may be written in terms of 
$\alpha$ while on the R.H.S. we can eliminate $v(y_0)$ completely  in terms of 
$\gamma_m^{\rm var}$ by making use of Eq.~(\ref{eq: gamma_m in v}).
One is 
left with a simple implicit algebraic equation for the anomalous dimension
\begin{equation}
\frac{3}{4} \alpha \>=\>  \left ( \, 1 + \gamma_m^{\rm var} \, \right ) \,  
\tan \left ( 
\frac{\pi/2  \; \gamma_m^{\rm var}}{1 + \gamma_m^{\rm var}} \right ) \> ,
\label{impl eq for gamma_m(alpha)}
\end{equation}
without ever having actually solved the variational equations themselves.
Eq. (\ref{impl eq for gamma_m(alpha)}) is the main result of this paper.

\section{Discussion}

\begin{figure}[htb]
\begin{center}
\epsfig{angle=90,height=8cm,file=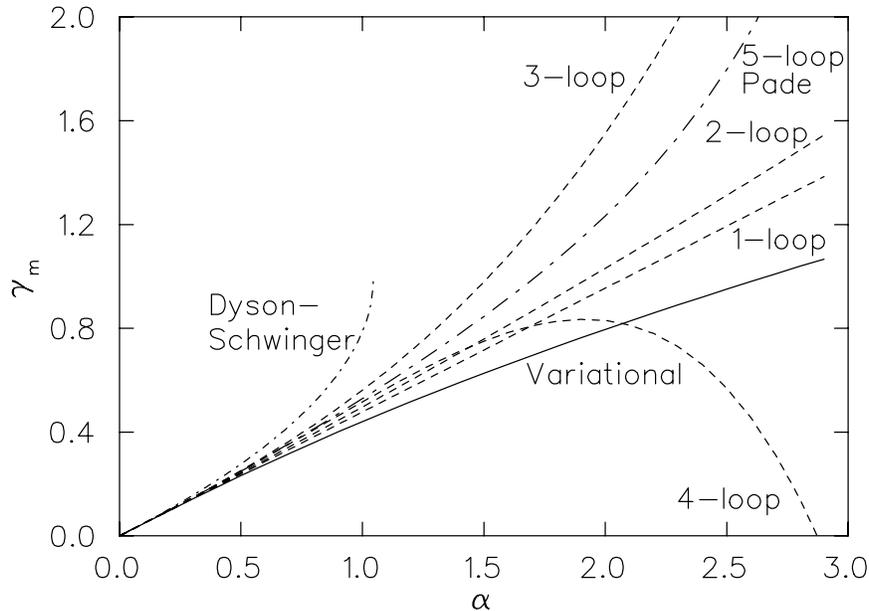}
\end{center}
\caption{Anomalous mass dimension $\gamma_m$ as function of the
coupling constant $\alpha$ in quenched QED. The variational result
(\ref{impl eq for gamma_m(alpha)}) is shown as a solid curve while the
solution from the Dyson-Schwinger equations in rainbow approximation
is indicated as a dot-dashed curve.  The curves labeled ``n-loop''
show the result up to n-loop perturbation theory.  Finally, the Pad\'e
estimation of the 5-loop result is also shown.}
\label{fig: 1}
  \vspace{0.5cm}
\end{figure}

When expanded in powers of $\alpha$, Eq.~(\ref{impl eq for gamma_m(alpha)}) 
immediately yields
\begin{equation}
\gamma_m^{\rm var}(\alpha) \>=\> \frac{3}{2} \frac{\alpha}{\pi} - 
\frac{9}{32} \pi^2 \left (\frac{\alpha}{\pi} \right )^3  
+ \frac{27}{32} \pi^2 \left (\frac{\alpha}{\pi} \right )^4 
- \frac{243}{128} \pi^2 \left ( 1 -
\frac{\pi^2}{20} \right )   \left (\frac{\alpha}{\pi} \right )^5 
+ {\cal O} (\alpha^6) \> ,
\label{gammavar pert}
\end{equation}
which may be compared to perturbation theory.  Numerically the values
of the coefficients are different but of the same order of magnitude
as the exact perturbative results. Note, however, that this comparison
is not particularly meaningful: the variational result is an
approximation which is valid at all $\alpha$. It need not have the
same, or even approximately the same, perturbative expansion in
$\alpha$ as the exact result. It should, however, be {\it numerically}
similar.  In Fig.~\ref{fig: 1} we plot the variational result as a function of
the coupling and compare it to perturbation theory up to 4-loop
order. For $ \alpha$ \raisebox{-1mm}{$\stackrel{>}{\sim}$} $ 1 $ the
3- and 4-loop anomalous dimensions start to deviate so much from each
other that one cannot trust either of them.  Also shown is the result
up to 5 loops, where the 5-loop coefficient has been estimated from
Pad\'e approximations to the perturbation theory (see
Eq. (2.12) of Ref. \cite{Pade}, which needs to be adapted to QED with
$n_f = 0 $ flavours; one finds $ \> \gamma_4^{\rm Pade} \>=\> 3.848 \>
$).  Clearly this does not significantly extend the numerical validity
of the perturbative result.  In short, the variational estimate for
$\gamma_m$ is roughly in agreement with (albeit apparently a little
below) the perturbative result in the region where the perturbative
result can be trusted.

Also shown in Fig.~\ref{fig: 1} is the only other easily available
non-perturbative result for $\gamma_m^{\rm MS}$ based on the use of
dimensionally regularized Dyson-Schwinger (DS) equations in ``rainbow
approximation'' within the Landau gauge.  This may be
obtained by adapting the discussion in Ref.\cite{chiSB} to finite $M_0$,
 with the
result that $\gamma_m^{\rm DS} = 1 - \sqrt{1 - 3 \alpha/\pi} $ (which
is the same as derived by Miransky \cite{Mir} using a hard momentum
cutoff).  We see that this result deviates from perturbation theory in
a region where, at least numerically, perturbation theory still
appears to converge.  Above $\alpha = \pi/3 = 1.047$ the DS result
becomes complex, this value of the coupling constant coinciding with
the coupling $\alpha_{cr}$ at which the onset of chiral symmetry
breaking takes place in those calculations.  This is in contrast
to the variational result which remains real for all values of the
coupling and in fact has the strong coupling limit
\begin{equation}
\gamma_m^{\rm var} (\alpha) \> \stackrel{\alpha \to \infty}{\longrightarrow} 
\> \frac{1}{4} \, \sqrt{6 \pi \alpha} - \frac{1}{2} + {\cal O} \left ( 
\frac{1}{\sqrt{\alpha}} \right ) \>.
\label{eq: large alpha}
\end{equation}

Further investigations are necessary to clarify the absence of any
obvious sign of chiral symmetry breaking in the variational result for
$\gamma_m^{\rm MS}(\alpha)$~\footnote{The reader should note that the issue of dynamical chiral 
symmetry breaking 
in a dimensionally regulated theory is a notoriously subtle problem; see
Ref.~\cite{chiSB}.  In particular, it was shown there that if four 
dimensional quenched QED breaks 
chiral symmetry {\it above} a critical coupling than 
the dimensionally regularized
theory will break it for {\it all} 
couplings at finite $\epsilon$.}.  
Indeed, in order to investigate the issue of dynamical chiral
symmetry breaking, 
it would seem to be more straightforward, 
at least
conceptually, to set $M_0$ on the right hand side of Mano's equation 
(\ref{Mano eq})
to zero and to see if the variational equations can be satisfied in this
case (for a finite physical mass $M$).  This, however, goes considerably 
beyond the scope of this
paper:  we have merely calculated $Z_M^{\rm MS}=M_0/M_\nu$ (or, more 
precisely, 
$\gamma_m^{\rm MS}$), which meant that we could simplify the calculation by 
{\it i}) restricting ourselves to considering the supersymmetric massless 
limit of 
the variational equations in Section II and {\it ii}) only taking into
account the most divergent contributions (as $\epsilon \rightarrow 0$) to
the variational equations, as well as to $W_2$, in Section III and 
the Appendix of this paper.  In a full calculation of the R.H.S. of Mano's 
equation (i.e. the additional calculation of the finite renormalization 
$M_\nu/M$) these two simplifications should not be made~\footnote{This 
situation is analogous to what is the case in perturbative 
calculations, where anomalous dimensions of operators are far easier to 
calculate than finite contributions.}.  In other words, even if $Z_M^{\rm MS}
\ne 0$, dynamical chiral symmetry breaking can still occur if $M_\nu/M$
vanishes for finite $M$.

It is interesting to note, however, that there are also some strong 
similarities
in the analytic structure of the variational and DS result. A perturbative 
inversion of
Eq.~(\ref{impl eq for gamma_m(alpha)}) , i.e. the expansion
$\gamma_m^{\rm var}(\alpha) = \sum_{n=1}^\infty c_n \alpha^n$, has a
finite radius of convergence due to a branch cut in the complex $\alpha$ plane.  
The position of this cut, and hence the radius of convergence, can be determined
most easily by searching for the value of $\alpha$ at which 
Eq.~(\ref{impl eq for gamma_m(alpha)})
has two solutions for $\gamma_m^{\rm var}$ which are infinitesimally close to 
each other.  This amounts to demanding that 
Eq.~(\ref{impl eq for gamma_m(alpha)}) is satisfied and at the same time
the derivative of its R.H.S. vanishes, i.e.
\begin{equation}
0\>=\> \cot \left ({\pi/2 \over 1 + \gamma_m^{\rm var}}\right )
\> + \> {\pi/2 \over 1 + \gamma_m^{\rm var}}\left / \sin^2 
\left ({\pi/2 \over 1 + \gamma_m^{\rm var}}\right )\right. \;\;\;.
\end{equation}
One finds that  $ \alpha_{con} = 0.7934$, which is not too different from 
the radius of convergence of the DS result~\cite{remark7}.  It is not clear whether
this similarity between $\alpha_{cr}$ and $\alpha_{con}$ is accidental
or not.  

In connection with this, it is interesting to note that for
large $n$ the behaviour of the expansion coefficients $c_n$ in both
the variational result as well as the DS result are rather similar:
\begin{equation}
c_n \> \approx \> \alpha_{con}^{-n}\> \frac{e^{- \beta}}{n^{3/2}}
\sin \left [ \left (a+{5 \pi \over 7}\right ) n - {3 \pi \over 7} + b \right ] \;\;\;,
\end{equation}
where numerically $\beta \approx 1.38$, $a \approx 2.3 \times
10^{-3}$ and $b \approx -8.27 \times
10^{-2}$. For the
DS result one obtains $\beta = \log( 2 \sqrt{\pi}) = 1.27$ and the sine
function is absent.  It is the
sine function in the variational result which is responsible for placing
the branchpoint (which, for the DS result,
is on the positive real axis) into the complex plane.  Furthermore, it is 
remarkable that the large-$\alpha$ limit
of $|\gamma_m(\alpha)|$ obtained in Eq.~(\ref{eq: large alpha}) is almost the 
same as for the DS result: 
$|\gamma_m^{\rm var}(\alpha)| \rightarrow 1.09 \sqrt{\alpha}\> $ vs. 
$\> |\gamma_m^{\rm DS}(\alpha)| \rightarrow 0.98 \sqrt{\alpha} $.

It should be pointed out that a finite radius of convergence of the perturbation
expansion is {\it not} what one generally expects from calculations of large 
orders of perturbation theory using
the methods of Lipatov and others~\cite{large orders}.  Rather, the 
factorial growth of the number of diagrams at n$^{th}$ order in perturbation 
theory tends to lead
to a vanishing radius of convergence.  As has been observed
elsewhere~\cite{WC}, it
can be shown that the variational calculation contains (pieces of) all possible
Feynman diagrams  at any order in perturbation theory.  One concludes, therefore, that
at n$^{th}$ order in perturbation theory there are either strong cancellations 
between diagrams in the variational calculation  or that ${\cal O}(n!)$ of them
give a vanishing contribution.

\section{Summary and Outlook}
 
We have applied polaron variational
techniques to quenched QED in $3+1$ dimensions and obtained, within
the MS scheme, a remarkably simple expression for the
anomalous mass dimension valid for arbitrary couplings. The approach
has considerable advantages over other techniques in that it
automatically maintains gauge invariance, as well as the requirements
of the renormalization group, and corrections can be systematically
calculated (as has been done in the polaron case
\cite{corr}). Furthermore, we have shown that the numerical results
for $\gamma_m$ are rather reasonable at small coupling and that at
large couplings the perturbative expansion of this quantity fails in a
way similar to rainbow DS results.  It would be interesting to compare
to DS calculations which go beyond the ladder approximation, thus
decreasing the strong gauge dependence inherent in that
approximation.  Furthermore, variational calculations with more
general trial actions could give an indication whether this analytic structure
is robust, thus indicating possible large cancellations between diagrams 
at high
order in the perturbation theory of quenched QED, or whether this structure
 is just
an artifact of the particular trial action used in this paper.
Finally, we note
that the calculation of physical observables or application to bound
state problems also seem feasible within the variational worldline approach
developed here.

\begin{acknowledgements}
We would like to thank Reinhard Alkofer for helpful discussions.
One of us (AWS) is supported by the Australian Research 
Council through an Australian Research Fellowship.
C.A. would like to thank PSI for its hospitality on several visits
during which  parts of this work were done.
\end{acknowledgements}

\section*{Appendix}
\appendix
\renewcommand{\theequation}{A.\arabic{equation}}
\setcounter{equation}{0}

In this Appendix we provide some of the technical details which enter
into the derivation of the variational approximation to the anomalous
dimension.  To begin with, we shall scale the trivial $\sigma$ dependence
out of $\mu^2(\sigma)$ and define the reduced pseudotime $s(\sigma)$ as
\begin{equation}
\mu^2(\sigma) \> = \> \sigma \> s(\sigma)\;\;\;.
\end{equation}
As argued in the main text, mass terms can be dropped for the calculation of
the mass anomalous dimension. A perturbative evaluation of the variational 
equations (\ref{var eq for A(E)}) and (\ref{def pseudotime}) for $ M = 0 $ 
then shows that profile function and reduced pseudotime have an expansion 
in powers of $E^{-\epsilon}$ and $\sigma^{\epsilon}$, respectively:
\begin{equation} 
A(E) \> = \>  1 + \sum_{n=1} A_n \left ( \frac{\nu^2}{E} 
\right )^{n \epsilon} \> \> \> ,\> 
s (\sigma) \> = \>   1 + \sum_{n=1} s_n \, \left ( \nu^2 \sigma 
\right )^{n \epsilon} \> .
\label{A,s perturb}
\end{equation}
One finds 
\begin{eqnarray}
A_1 &=& c_{\epsilon} \Gamma(\epsilon) \cos \left ( \frac{\epsilon \pi}{2}
\right ) \> \stackrel{\epsilon \to 0}{\longrightarrow} \> 
\frac{3 \alpha}{2 \pi} \frac{1}{\epsilon} \> \> ,\> \>
s_1  \> = \>  - \frac{c_{\epsilon}}{\epsilon (1 + \epsilon)} 
\>  \stackrel{\epsilon \to 0}{\longrightarrow} \>
 - \frac{3 \alpha}{2 \pi} \frac{1}{\epsilon} \nonumber \\
A_2 &=& \frac{1}{2} \left ( \frac{c_{\epsilon}}{\epsilon} \right )^2 \,
\frac{1-\epsilon}{1+\epsilon} \, \Gamma(1+2 \epsilon) \, \cos ( \epsilon \pi )
\> \stackrel{\epsilon \to 0}{\longrightarrow} \>
\frac{1}{2} \left ( \, \frac{3 \alpha}{2 \pi} \frac{1}{\epsilon} \, \right )^2
 \\
s_2 &=& \frac{1}{2} \left ( \frac{c_{\epsilon}}{\epsilon} \right )^2 \,
\frac{1}{1+2 \epsilon} \, 
\left [ \, \left ( 1 + \frac{1}{\cos (\epsilon \pi)} \right ) \,
\frac{\Gamma^2(1+\epsilon)}{\Gamma(1+2 \epsilon)} - 
\frac{1-\epsilon}{1+\epsilon} \, \right ] 
\> \stackrel{\epsilon \to 0}{\longrightarrow} \>
\frac{1}{2} \left ( \, \frac{3 \alpha}{2 \pi} \frac{1}{\epsilon} \, \right )^2
\;\;.
\nonumber
\end{eqnarray}
This suggests that perhaps the leading $\epsilon$-behaviour of the 
coefficients 
is
\begin{equation}
A_n  \> \stackrel{\epsilon \to 0}{\longrightarrow} \> 
\frac{1}{n!} \left ( \, \frac{3 \alpha}{2 \pi} \frac{1}{\epsilon} \, 
\right )^n \> , \> \> 
s_n \> \stackrel{\epsilon \to 0}{\longrightarrow} \>  
 \frac{(-1)^n}{n!} \left ( \, \frac{3 \alpha}{2 \pi} \frac{1}{\epsilon} \,
\right )^n\;\;\;.
\label{an,sn leading eps}
\end{equation}
In view of these results we rewrite all equations in terms of the 
dimensionless quantities
\begin{equation}
y \> = \> c_{\epsilon} \left (\nu^2 \sigma \right )^{\epsilon} \> \> ,\>
z \> = \>  c_{\epsilon} \left ( \frac{\nu^2}{E} \right )^{\epsilon}
\end{equation}
where we have also rescaled by $c_\epsilon$ (which is linear in the coupling)
because the mass scale $\nu$ always appears in the combination 
$\alpha \, \nu^{2 \epsilon}$ in dimensional regularization.
In a similar way, the variational parameter $\lambda$ almost always appears
in the combination $\lambda M$, so it is convenient to define the
dimensionless combination
\begin{equation}
a_{\epsilon} \> = \> c_\epsilon \,
\left ( {2 \kappa_0 \nu^2 \over i \lambda^2 M^2} 
\right)^\epsilon \;\;\;.
\label{def a_epsilon}
\end{equation}

With these definitions, the massless variational equation 
(\ref{var eq for A(E)}) for $A(E)$ 
may be brought into the form
\begin{equation}
A(z) \> = \> 1 \> + \> {1 \over \epsilon} \int_0^\infty
dy \> {\cos (y/z)^{1/\epsilon} \over [s(y)]^{1-\epsilon} }\;\;\;.
\label{eq: scaled a}
\end{equation}
The reduced pseudotime [ i.e. the rescaled version of
Eq.~(\ref{def pseudotime}) ] is now given by
\begin{equation}
s(y) \> = \> {2 \over \pi} {1 \over \epsilon} \int_0^\infty
dz \> {1 \over z} \left ({z \over y}\right)^{1/\epsilon}
{1 - \cos  ( y /z )^{1/\epsilon} \over A(z)}\;\;\;,
\label{eq: scaled s}
\end{equation} 
while the rescaled potential  $W_2$ becomes a function of $a_{\epsilon}$ alone:
\begin{equation}
W_2(a_{\epsilon})\>=\> {(2-\epsilon)(1-\epsilon)\over 2 \epsilon}
\int_0^\infty {dy  \over [s(y)]^{1-\epsilon}}
\int_0^1 {du \over u^\epsilon} \>
[\epsilon+(1-\epsilon)\> u] \>
\exp\left [ - (y/a_{\epsilon})^{1/\epsilon} {u \over s(y)} \right ]
\label{eq: scaled w2}
\end{equation}
and hence the variational equation (\ref{eq: var eq for lambda}) for $\lambda$ 
becomes
\begin{equation}
{1 \over \lambda} \> = \> 1 \> + \> 2  W_2(a_{\epsilon})
\> - \> 2 \epsilon a_{\epsilon} W_2'(a_{\epsilon})\;\;\;.
\label{scaled var eq for lambda}
\end{equation}

The anomalous mass dimension in the MS scheme may be defined 
[ see Eq.~(\ref{gamma_m def}) ] through
\begin{equation}
\gamma_m \> = \> - \lim_{\epsilon \rightarrow 0}
\epsilon \> {\alpha \over Z_M^2}\> {\partial \over \partial \alpha}  Z_M^2\;\;\;.
\end{equation}
Note that this equation is correct independently
of whether $Z_M$ has been calculated in the MS scheme or whether
it is defined through Mano's equation by $Z_M \> \equiv \> M_0/M$.
It is because of this fact that we can derive a nonperturbative expression
for $\gamma_m$ in the MS scheme, even though this scheme is usually
only used within the context of perturbation theory.

As the $\alpha$ dependence of $Z_M$ now only enters through the variable 
$a_{\epsilon}$ (and of course implicitly through the variational parameters), 
it is not surprising that we may use
the variational equation (\ref{scaled var eq for lambda}) for $\lambda$ 
to simplify $\gamma_m$.  Indeed, differentiating Mano's equation with respect
to the coupling gives
\begin{equation}
{\partial \over \partial \alpha} Z_M^2 \> = \>
{\partial\lambda \over \partial \alpha}{\partial \over \partial \lambda} Z_M^2 
\> - \> 2 {\partial a_{\epsilon} \over \partial \alpha} 
\lambda^2 W_2'(a_{\epsilon})\;\;\;.
\end{equation}
The first term is zero because of the variational equation for $\lambda$,
$\partial a_{\epsilon} /\partial \alpha $ is just 
$a_{\epsilon}/ \alpha$ and by substituting the variational equation for 
$\lambda$ into Mano's equation we find
\begin{equation}
Z_M^2 \> = \> \lambda \, [ \, 1 - 2 \epsilon \lambda a_{\epsilon} 
W_2'(a_{\epsilon}) \, ] \> .
\end{equation}
Hence the anomalous mass dimension is just given by
\begin{equation}
\gamma_m \> = \> \lim_{\epsilon \rightarrow 0} \,  
{2 \epsilon \lambda a_{\epsilon} W_2'(a_{\epsilon}) \over
1 \> - \> 2 \epsilon \lambda a_{\epsilon} W_2'(a_{\epsilon})}\;\;\;.
\label{eq: gamma_m app}
\end{equation}

In order to proceed further, we need to evaluate $W_2(a_{\epsilon})$.  In
general one would need to do this numerically, however fortunately
in Eq.~(\ref{eq: gamma_m app}) only the  small-$\epsilon$ limit
is required.  Let us assume that the reduced pseudotime may be written as
\begin{equation}
s(y) \> = \> \exp \left [ - \frac{\omega(y,\epsilon)}{\epsilon} \right ]\;\;,
\label{eq: omega def}
\end{equation}
where  $ \omega_0(y) \equiv \lim_{\epsilon \rightarrow 0} \omega(y,\epsilon) $
 is finite. This is supported by
the perturbative results given in Eqs. (\ref{A,s perturb}) and  
(\ref{an,sn leading eps}) and 
we shall show that this holds in general when we solve the
variational equations below.  In this case the exponential
in Eq.~(\ref{eq: scaled w2}) has the argument
\begin{equation}
-\left ({y \over a_{\epsilon}} e^{\omega(y,\epsilon)}\right )^{1/\epsilon} u\;\;\;.
\end{equation}
If the term in brackets is
larger than one, this argument will
become arbitrarily large (and negative) in the limit $\epsilon \rightarrow 0$,
hence it will lead to a vanishing contribution to the integral.  If
the term in brackets is smaller than one, however, the argument goes to
zero, the exponential factor in  Eq.~(\ref{eq: scaled w2}) may be replaced by
unity and the integral over $u$ may be performed, yielding
\begin{equation}
W_2(a_{\epsilon})\>\stackrel{\epsilon \rightarrow 0}{\longrightarrow}\> 
{1 \over 2 \epsilon}
\int_0^{y_0} dy \> \exp \left[{\omega(y,\epsilon) \over \epsilon} 
(1-\epsilon)\right ]\;\;\;,
\label{eq: w2 approx int}
\end{equation}
where $y_0$ is given by the equation
\begin{equation}
y_0 \, e^{\omega_0(y_0)} \> = \>  \lim_{\epsilon \to 0} a_{\epsilon} \> .
\label{eq: y_0 def}
\end{equation}
We have assumed here that $y e^{\omega_0(y)}$ is an increasing 
function of $y$, which will turn out to be the case.
The leading term in Eq.~(\ref{eq: w2 approx int}) may be obtained by
integration by parts, with the result
\begin{equation}
W_2(a_{\epsilon}) \> \stackrel{\epsilon \rightarrow 0}{\longrightarrow} \> 
{1 \over 2 \omega'(y_0,\epsilon)} \exp  
\left[{\omega(y_0,\epsilon) \over \epsilon} (1-\epsilon)\right ]\;\;\;.
\end{equation}
We also require the derivative of this function, which is most
easily obtained by direct differentiation of Eq.~(\ref{eq: w2 approx int}):
\begin{equation}
W_2'(a_{\epsilon}) \> \stackrel{\epsilon \to 0}{\longrightarrow} \> 
{1 \over 2 \epsilon} \exp  
\left[{\omega(y_0,\epsilon) \over \epsilon} (1-\epsilon)\right ]
{e^{-\omega(y_0,\epsilon)} \over 1 \> + \> y_0 \omega'(y_0,\epsilon)}\;\;\;.
\end{equation}
Substitution into the variational equation for $\lambda$ yields
\begin{equation}
\lambda \> \stackrel{\epsilon \to 0}{\longrightarrow} \> 
\omega'(y_0,\epsilon) \,  
{1\> + \> y_0 \, \omega'(y_0,\epsilon) \over \exp  
\left[{\omega(y_0,\epsilon) \over \epsilon} (1-\epsilon)\right ]}
\label{lambda}
\end{equation}
and hence 
\begin{equation}
2 \epsilon \lambda a_{\epsilon} W_2'(a_{\epsilon}) \> 
\stackrel{\epsilon \to 0}{\longrightarrow} \> y_0 \, \omega'(y_0,\epsilon)
\end{equation}
so that the anomalous dimension becomes
\begin{equation}
\gamma_m \> = \> { y_0 \, \omega'_0(y_0) \over 1 -  y_0 \omega'_0(y_0)}\;\;\;.
\end{equation}
We stress that only the last line is exact while the previous
ones have correction terms for finite $\epsilon$.  In particular,
the calculation of $Z_M = M_0/M$ (as opposed to $Z_M^{\rm MS} = M_0/M_\nu$)  would 
require these additional
terms and hence the result, unfortunately, does not shed light
on whether $Z_M$ could in fact be zero for finite $M$, which would
signal chiral symmetry breaking. Note that the limit $ \epsilon \to 0 $
in Eq. (\ref{eq: y_0 def}) needs some care: naively, one 
would conclude from the definition (\ref{def a_epsilon}) that the  R.H.S. 
equals $\lim_{\epsilon \to 0} c_{\epsilon} = 3 \alpha/(2 \pi) $ but
Eq. (\ref{lambda}) shows that the variational parameter $\lambda$ vanishes
like $ \exp(-\omega_0(y_0)/\epsilon) $ and therefore also gives a contribution:
\begin{equation}
y_0 \, e^{\omega_0(y_0)} \> = \> \frac{3 \alpha}{2 \pi} \,  
e^{2 \omega_0(y_0)} \;\;\;,
\end{equation}
this being the result (\ref{implicit}) quoted in the main text.

It now remains to calculate the function $\omega_0(y)$.  The arguments
used to derive the approximate expression for $W_2(a_{\epsilon})$ in 
Eq.~(\ref{eq: w2 approx int})
are more difficult to apply to the variational equation (\ref{eq: scaled a}) 
for $A(z)$  and
the definition (\ref{eq: scaled s}) of $s(y)$ because of the rapidly 
oscillating trigonometric functions appearing in their integrands.
We shall therefore adopt a more systematic approach at this stage and
note that it is possible to write these equations in a
differential form. Consider, for example, an integral of the type
\begin{equation}
I_{\epsilon}(z) \> =\> \int_0^{\infty} dy \> f(y) \, 
\cos \left (\frac{y}{z} \right )^{1/\epsilon}\;\;\;. 
\label{def I}
\end{equation}
By changing integration variable to $y^{1/\epsilon}$ and Taylor expanding
the function $f(y)$ we may carry out the integration term by term by making
use of the integral
\begin{equation}
\int_0^{\infty} dy \> y^{q-1} \, \cos  y \> = \>
\Gamma(q) \, \cos \left (q \frac{\pi}{2} \right ) \;\;\;.
\label{eq: y int}
\end{equation}
Hence we obtain
\begin{equation}
I_{\epsilon}(z) \> =\>  \sum_{n=0}^{\infty} f^{(n)}(0) \, \frac{z^{n+1}}{(n+1)!}
\> \Gamma \left [ 1 + (n+1) \epsilon \right ] \, \cos \left [ (n+1) \epsilon 
\frac{\pi}{2} \right ] \;\;\;.
\label{I syst}
\end{equation}
This expression may be resummed, by defining the dilatation 
operator $D_z \equiv z {d \over dz}$, into the compact form
\begin{equation}
I_{\epsilon}(z) \> = \> 
\Gamma (1 + \epsilon D_z) \, \cos \left ( \frac{\pi}{2} \epsilon D_z
\right ) \, \int_0^z dy \, f(y) \> =: \> \gamma_c \left (\epsilon D_z \right )
\, \int_0^z dy \, f(y) \> . 
\end{equation}
Hence the variational equation for $A(z)$ becomes
\begin{equation}
A (z) \>=\> 1 + \frac{1}{\epsilon} \, 
\gamma_c(\epsilon D_{z}) \int_0^{z}
dy \> \frac{1}{[s(y)]^{1-\epsilon}} 
\label{eq: z,y var eq2} 
\end{equation}
and in a similar way we can rewrite Eq.~(\ref{eq: scaled s}) as
\begin{equation}
s(y) \> =\>   
\frac{1}{1+\epsilon D_{y}} \> \frac{1}{\gamma_c(\epsilon D_{y})} \> 
\frac{1}{A(y)}\;\;\;.
\label{eq: z,y var eq3}
\end{equation}
Inverting Eq.~(\ref{eq: z,y var eq3}) and substituting into 
Eq.~(\ref{eq: z,y var eq2})
eliminates the profile function $A(z)$:
\begin{equation}
\frac{1}{ \left (1 + \epsilon D_{y} \right)
\gamma_c (\epsilon D_{y}) \, s(y) }
\> = \>  1 + 
\frac{1}{\epsilon} \gamma_c (\epsilon D_{y}) \int_0^{y} dx \>  
\frac{1}{[s(x)]^{1-\epsilon}} \> \;\;.
\end{equation}
Finally, one can eliminate the integral by operating with $D_y$ on 
both sides of this equation, so that
\begin{equation}
D_{y} \frac{1}{(1+\epsilon D_{y})
\gamma_c (\epsilon D_{y}) \, s(y)} \>=\> \frac{1}{\epsilon}
\gamma_c (\epsilon D_{y}) \, \frac{y}{[s(y)]^{1-\epsilon}} \> .
\label{nonlinear eq for s}
\end{equation}
This equation may be solved systematically by defining $s(y)$ in
terms of the function $\omega(y,\epsilon)$ [ see Eq.~(\ref{eq: omega def}) ]
and by making the Ansatz that  $\omega(y,\epsilon)$ has
a power expansion in $\epsilon$
\begin{equation}
\omega(y,\epsilon) \> = \> \omega_0(y) \> + \> \epsilon \; \omega_1(y)
 \> + \> \ldots\;\;\;.
\label{eq: ansatz}
\end{equation}
The crucial observation is that repeated application of the dilatation
operator on an exponential of the form of Eq.~(\ref{eq: omega def})
results in 
$ \left ( \epsilon D_{y} \right )^n \exp \left ( \omega / \epsilon \right ) 
=  \left [ (y \,  \omega')^n  + {\cal O} (\epsilon) \right ]
\, \exp \left (\omega/\epsilon \right ) $
so that at leading order in $\epsilon$, for any
function $F(\epsilon D_{y})$ acting on $ \exp(\pm\omega/\epsilon) $,
we have
\begin{equation}
F(\epsilon D_{y}) \, \exp(\pm \omega/\epsilon) \> 
\stackrel{\epsilon \to 0}{\longrightarrow} \>  F(\pm y \, \omega_0')
\, \exp(\pm \omega/\epsilon) \> .
\end{equation}
Applying this relation to Eq.~(\ref{nonlinear eq for s}) provides the 
following equation for $\omega_0(y)$
\begin{equation}
\frac{\omega_0'}{(1-y \, \omega_0') \, \gamma_c(-y \, \omega_0')} 
\> = \> \gamma_c(y \, \omega_0') \, e^{- \omega_0} \> .
\label{diff eq0 for omega0}
\end{equation}
which is $\epsilon$-independent, justifying
the Ansatz~(\ref{eq: ansatz}) {\it a postiori}. This equation may
be simplified considerably by making use of the reflection formula
 $ \Gamma(z) \, \Gamma(1-z) =   \pi/\sin \pi z $ for $\Gamma$-functions.
By defining $v(y) \equiv y\; \omega_0'(y)$, we then find 
\begin{equation}
{e^{\omega_0(y)} \over y} \> = \> {\pi \over 2} [1-v(y)] \cot \left [ 
{\pi \over 2}v(y) \right ]\;\;\;,
\label{diff eq}
\end{equation}
which is Eq. (\ref{eq: final var eq}) in the main text.
Together with the boundary condition $\omega_0(0) = 0$ [ i.e. 
$\mu^2(\sigma) \rightarrow \sigma $ for $\sigma \rightarrow 0$,
as discussed in the footnote below Eq.~(\ref{def pseudotime}) ] the first-order
nonlinear differential equation (\ref{diff eq})  
determines the function $\omega_0(y)$.  Remarkably, as shown in the 
main text, it is not actually necessary to solve it in order to obtain 
the anomalous mass dimension $\gamma_m$.
It is also interesting to note that due to the reflection formula
all $\Gamma$-functions have disappeared, 
which has the consequence that in a perturbative expansion of 
$\gamma_m^{\rm var}$ no Riemann $\zeta$-functions, but only powers of $\pi$, 
occur.


\begin{thebibliography}{99}
\bibitem{remark5} If the ultraviolet divergences are regulated by a formfactor
or a hard cutoff, dimensional arguments show 
that $V_2$ is suppressed by a factor ($M$/cutoff)$^2$ compared to $V_1$.
\bibitem{remark7} The reader should note that DS calculations which go
beyond the rainbow approximation tend to reduce the value of the critical
coupling to about $\alpha_{cr} \approx 0.9$, which is even closer
to the radius of convergence of the variational result.
\end{thebibliography}
\end{document}